\documentclass[twocolumn,aps,prl,superscriptaddress,showpacs,byrevtex]{revtex4-1}
\usepackage{epsf,graphicx,amssymb}
\usepackage[usenames]{color}
\usepackage{natbib}
\usepackage{float,ulem}
\usepackage{gensymb}
\usepackage{amsmath,amssymb}
\usepackage{color}
\begin{document}

\title{Saturation of the inverse cascade in surface gravity wave turbulence}

\author{E. Falcon}
\email[E-mail: ]{eric.falcon@univ-paris-diderot.fr}
\affiliation{Universit\'e de Paris, Univ Paris Diderot, MSC, UMR 7057 CNRS, F-75 013 Paris, France}
\author{G. Michel}
\affiliation{Sorbonne Universit\'e, IJLRA, UMR 7190 CNRS, F-75 005 Paris, France}
\author{G. Prabhudesai}
\affiliation{Ecole Normale Sup\'erieure, LPS, UMR 8550 CNRS, F-75 205 Paris, France}
\author{A. Cazaubiel}
\affiliation{Universit\'e de Paris, Univ Paris Diderot, MSC, UMR 7057 CNRS, F-75 013 Paris, France}
\author{M. Berhanu}
\affiliation{Universit\'e de Paris, Univ Paris Diderot, MSC, UMR 7057 CNRS, F-75 013 Paris, France}
\author{N. Mordant}
\affiliation{Université Grenoble Alpes, LEGI, UMR 5519 CNRS, F-38 000 Grenoble, France}
\author{S. Auma\^{\i}tre}
\affiliation{CEA-Saclay, Sphynx, DSM, URA 2464 CNRS, F-91 191 Gif-sur-Yvette, France}
\author{F. Bonnefoy}
\affiliation{Ecole Centrale de Nantes, LHEEA, UMR 6598 CNRS, F-44 321 Nantes, France}

\date{\today}

\begin{abstract}     
We report on the observation of surface gravity wave turbulence at scales larger than the forcing ones in a large basin. In addition to the downscale transfer usually reported in gravity wave turbulence, an upscale transfer is observed, interpreted as the inverse cascade of weak turbulence theory. A steady state is achieved when the inverse cascade reaches a scale in between the forcing wavelength and the basin size, but far from the latter. This inverse cascade saturation, which depends on the wave steepness, is probably due to the emergence of nonlinear dissipative structures such as sharp-crested waves.

\end{abstract}

\maketitle
\paragraph*{Introduction. \textemdash}

Wave turbulence is a phenomenon exhibited by random nonlinear waves in interaction. It occurs in various contexts: ocean surface waves, plasma waves, hydroelastic or elastic waves, internal waves, or optical waves \cite{Falcon2010revue,Zakharovbook,Nazarenkobook,Newell2011}. Nonlinear wave interactions are the basic mechanism that transfers the energy from a large (forcing) scale down to a small (dissipative) scale. Most experiments on wave turbulence concern its small-scale properties resulting from this direct energy cascade \cite{Shrira2013,Hawai}, notably to compare them with weak turbulence theory (WTT) \cite{Zakharovbook,Nazarenkobook}. The large-scale properties (i.e. larger than the forcing scale) have been much less investigated experimentally \cite{Deike2011,Bortolozzo2009,Ganshin2008,Michel2017}, although their understanding is of primary interest (e.g. for climate modeling and long-term weather forecasting).  For instance, for systems involving 4-wave interactions such as deep water gravity waves, an inverse cascade from the forcing scales towards a larger (dissipative) scale was predicted in the 80's by WTT, due to an additional conserved quantity (wave action)~\cite{Zakharov82}. It was confirmed by direct numerical simulations of \mbox{Zakharov} equations more than 10 years ago~\cite{Annenkov2006, Korotkevitch2008}. To our knowledge, the only laboratory observation of inverse cascade in gravity wave turbulence is limited to a narrow inertial range due to the small-size container used~\cite{Deike2011}, while recent attempts have been inconclusive within a larger size basin \cite{MordantEuromech2019} or using another type of forcing \cite{Nazarenko2016}.  As for field observations, inverse cascade is hardly distinguishable from the direct cascade due to anisotropy,  inhomogeneity, and nonstationnarity of the wind-generated wave fields \cite{Nazarenko2016,Romero2010,Oceano}. 
A downshifting of the spectrum peak is rather reported as the wind strength or fetch grows \cite{Romero2010,Kawai1979,Hara1991}. 



In this Letter, we report the first laboratory experiment in a large-scale basin evidencing the formation of an inverse cascade of gravity waves. To do this, it has been found necessary to replace the usual absorbing beach of large basins by a reflective wall, and to use a multidirectional forcing to foster wave interactions. The cascade is found to stop well before a so-called ``condensate'' state is reached, i.e., before wave action piles up at large scale due to basin finite size effects. Instead, a saturation is observed resulting from the emergence of highly dissipative nonlinear structures. Although well understood in 2D hydrodynamic turbulence \cite{Kraichnan67} or Bose-Einstein condensation \cite{Josserand05,Sun2012}, such large-scales dynamics resulting from an inverse cascade is far from being fully understood for wave turbulence systems beyond gravity surface waves \cite{Deike2011}, such as for optical waves \cite{Bortolozzo2009}, waves in superfluid \cite{Ganshin2008}, plasma waves \cite{Kats1971}, or elastic waves \cite{During2015,Hassaini2019}.

\paragraph*{Theoretical background.\textemdash}The dispersion relation of linear deep-water surface gravity waves is $\omega(k)=\sqrt{gk}$ with $\omega=2\pi f$ the angular frequency, $k$ the wave number, and $g$ the acceleration of gravity. For weakly nonlinear interacting waves in a stationary out-of-equilibrium state within an infinite size system, WTT predicts the spectrum of surface elevation $S_\eta$ for scales larger than the forcing one (inverse cascade) for a wave action flux $Q$ (per unit surface and density) \cite{Zakharov82}
\begin{equation}
S^{i}_{\eta}(f) \sim Q^{1/3}g f^{-11/3} {\rm \ ,}
\label{specinverse}
\end{equation}
or for an energy flux $P$ cascading from large to small scales (direct cascade) $S^{d}_{\eta}(f) \sim P^{1/3}g f^{-4}$ \cite{Zakharov67}. $S_{\eta}(f)$ has dimension $L^2T$, $P$ has dimension $(L/T)^3$, and $[Q]= [P]/ [\omega$]. WTT assumes a timescale separation between the nonlinear time, which reads for the inverse cascade $\tau^i_{nl} \sim Q^{-2/3}g^{1/6}k^{-11/6}$, and the linear timescale $\tau_{lin}=1/\omega$, that is the nonlinearity parameter $\tau_{lin}/\tau^i_{nl} \sim Q^{2/3}g^{-2/3}k^{4/3} \ll 1$ \cite{Newell2011}. As nonlinearity increases with $Q$, breaking of weak turbulence is expected to occur for $Q > Q_{c}=g/k^{2}$. 



\begin{figure}[!t]
\begin{center}
\includegraphics[width=8cm]{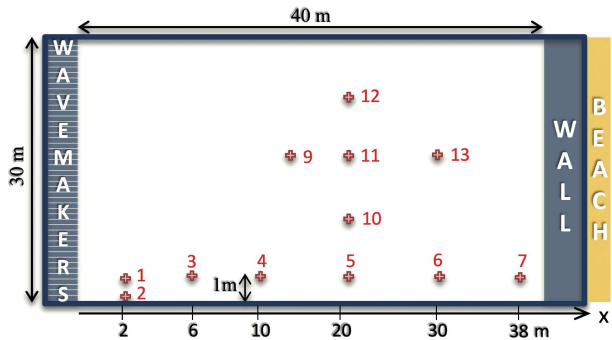}
\caption{Experimental setup showing the 48 flaps wave maker, the ending wall, and the locations of the 12 wave probes.} 
\label{fig01}
\end{center}
\end{figure} 

\paragraph*{Experimental setup.\textemdash}Experiments were performed in the large-scale wave basin (40 m long $\times  \ 30$ m wide $\times \ 5$ m deep) at Ecole Centrale de Nantes with a wave maker made of 48 independently controlled flaps located at one end of the basin (see Fig.\ \ref{fig01}). To favor homogeneity, wave reflections, and wave interactions necessary to observe an inverse cascade, a solid wall is built at the opposite end instead of the usual absorbing sloping beach. A multidirectional random wave forcing is generated around a central frequency $f_0=1.8$ Hz within a narrow spectral bandwidth $f_0\pm\Delta f$  with $\Delta f=0.2$ Hz. To be able to observe waves at lower frequencies than the forcing ones, $f_0$ is chosen to be close to the high-frequency limit of the wave maker. The 2D multidirectional forcing is also crucial to ensure a spatially homogeneous wave field, and to avoid the direct excitation of the first basin eigenmodes ($\sim 0.1$ -- 0.3 Hz). The surface elevation $\eta(t)$ is recorded during 1 hour by means of an array of 12 resistive wave probes located at different distances $x$ from the wave maker ($x=2$ to 38 m - see Fig.\ \ref{fig01}). Their vertical resolution is approximately 0.1 mm, their frequency resolution close to 20 Hz, and the sampling frequency 128 Hz. The typical rms wave height $\sigma_{\eta}\equiv \sqrt{\langle\eta^2(t)\rangle_t}$ is less than 2 cm, and the wave steepness $\epsilon \leq 0.1$ to limit to a weakly-nonlinear wave regime. The central wavelength generated is $\lambda_0=g/(2\pi f_0^2) \simeq 0.48$ m. The basin length is thus $83\lambda_0$ (assuming an unrealistic 1D monochromatic propagation), and a round-trip from the wave maker lasts $T \simeq 3$ min (group velocity $\omega_0/(2k_0)\simeq 0.44$ m/s).  

\begin{figure}[!t]
\begin{center}
\includegraphics[scale=0.46]{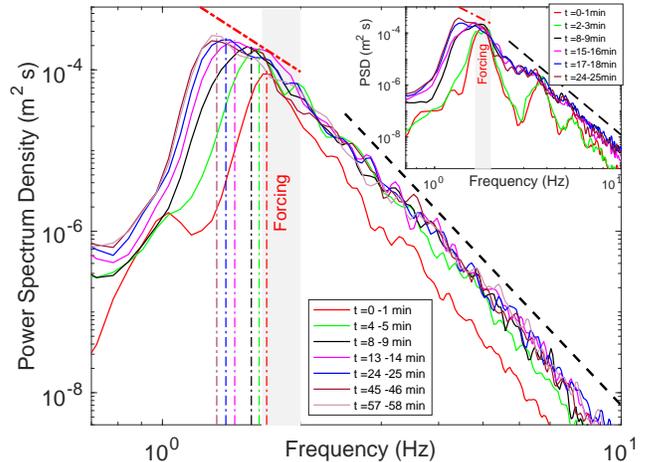}
\caption{Temporal evolution of the power spectrum density (PSD) of the wave height at $x= 20$ m (averaged on 4 probes n$^{\circ}$5, 10, 11, 12 - see Fig.\ \ref{fig01}). $\sigma_{\eta}=1.14$ cm. Gray area: Forcing bandwidth. Dash-dotted vertical lines correspond to the maximum of each spectrum. Dash-dotted red line is the prediction in $f^{-11/3}$ of Eq.\ (\ref{specinverse}).  Dashed black line displays a $f^{-6}$ fit. Inset: Same for $x=2$ m (averaged on probes n$^{\circ}$1, 2).} 
\label{fig02}
\end{center}
\end{figure}

\paragraph*{Wave spectrum.\textemdash}The temporal evolution of the power spectrum of the surface elevation recorded close to the wave maker ($x=2$ m) is shown in the inset of Fig. \ref{fig02}, the spectrum being computed over 1 min intervals. For $t \leq T$, the wave spectrum displays peaks at the forcing frequencies and at their harmonics, as expected. After typically more than a wave round trip ($t > T$), the main spectrum peak undergoes a shift towards low frequencies (or large scales), whereas frequency power-laws are observed at both low and high frequencies compared to the forcing ones. These two observations are related to the wave interaction dynamics occurring cumulatively after a while, and is not a wave maker's signature.  Note that the direct cascade towards high frequency differs significantly from WWT prediction in $f^{-4}$, as reported earlier in different basin sizes \cite{Falcon07,DeikeJFM2015,Aubourg2017}, this departure being ascribed to the modulation of bound waves \cite{Michel2018,CampagnePRF18}. Numerical simulations have also evidenced departures from this prediction resulting from the occurrence of an inverse cascade \cite{Korotkevitch2008}.  Concerning the large-scale transfer, the frequency downshifting of the main peak is observed homogeneously within the basin (see below). For instance, in the middle of the basin (see main Fig.\ \ref{fig02}), the spectral peak $f_{max}$ is shifted with time towards low frequencies and the spectral shape is compatible with the inverse cascade prediction of WTT in $f^{-11/3}$ of Eq.\ (\ref{specinverse}) (see dash-dotted red line). Although this experimental downshifting occurs in a narrow range ($1/6$ decade width in $f$) it corresponds to $1/3$ decade in $k$, a result similar to the ones achieved in numerical simulations of this inverse cascade \cite{Annenkov2006,Korotkevitch2008}. The saturation of this inverse cascade is also of interest: the wave spectrum reaches a stationary state in which the gravest eigenmodes are not excited and no condensate is observed. Since viscous dissipation is weak at these large scales (see below), it means that additional dissipation occurs to stop the inverse cascade.


\begin{figure}[!t]
\begin{center}
\includegraphics[scale=0.45]{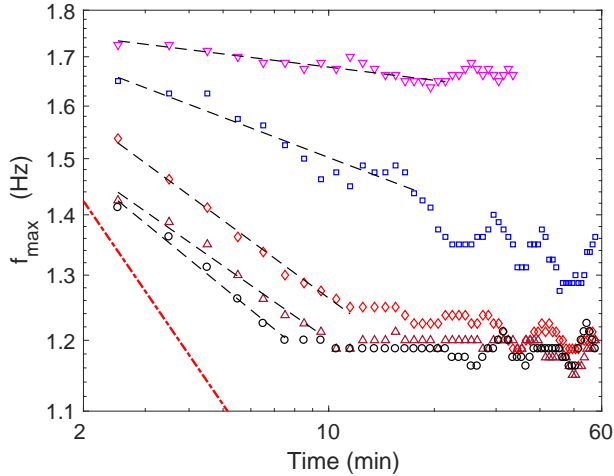}
\caption{Frequency of the spectrum maximum, $f_{max}$, over time for different forcing amplitudes. $\sigma^{sat}_{\eta}=$ ($\triangledown$) 0.59, ($\square$) 1.19, ($\lozenge$) 1.54, ($\vartriangle$) 1.83, and ($\circ$) 1.92 cm. Dash-dotted line is the prediction $t^{-3/11}$ \cite{Badulin2005}. Dashed lines are best power-law fits, at short times. $x=20$ m.}
\label{fig03}
\end{center}
\end{figure}

\begin{figure}[!t]
\begin{center}
\includegraphics[scale=0.45]{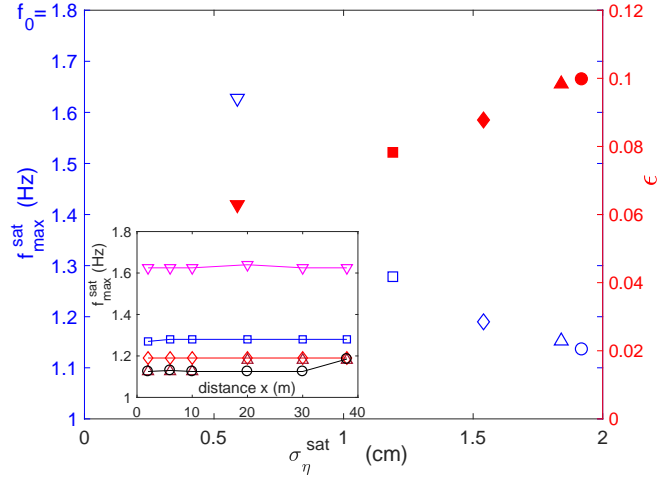}
\caption{Left-hand axis: Saturation frequency $f_{max}^{sat}$ reached by the spectrum versus forcing amplitude at $x=20$ m (blue open symbols). Right-hand axis: Wave steepness $\epsilon$ versus forcing at $x=20$ m (red full symbols). Inset: $f_{max}^{sat}$ for different distances $x$ along the basin and different forcing amplitudes. Same symbols as inset of Fig.\ \ref{fig03}.} 
\label{fig04}
\end{center}
\end{figure}

\paragraph*{Saturation.\textemdash}To better quantify this saturation, we report in Fig.\ \ref{fig03} the frequency of the spectrum maximum, $f_{max}$, versus time along with the WTT prediction $f_{max}\sim t^{-3/11}$ which assumes a linear growth of the total wave action $N \sim t^1$ \cite{Badulin2005,Pushkarev2003} (see dash-dotted red line). At short times, experimental values of $f_{max}$ are found to be of the form $t^{-\alpha}$ but $\alpha \in [0.1{\rm ,}\ 0.15]$ differs from the theoretical exponent ($-3/11= -0.273$). This departure could be ascribed to small scale dissipation not being negligible during this transient regime as well as the presence of a dual cascade, contrary to the assumption of WWT.
Because of the interaction between the wave maker and the wavefield, a constant wave action flux may also not be achieved (i.e. $N\sim \int Q dt\nsim t^1$).
Indeed, following \cite{Badulin2005,Pushkarev2003}, one finds that the total wave action growth should scale as $N \sim t^{(11\alpha-1)/2}$ that is between $t^{0.05}$ and $t^{0.35}$ using our values of $\alpha$. For long enough times, we observe that  $f_{max}$ saturates, and all the faster as the forcing amplitude increases (see Fig.\ \ref{fig03} - Long time oscillations of the spectral peak are related to spectrum resolution $\simeq 0.06$ Hz). This saturation frequency $f_{max}^{sat}$ decreases as the forcing amplitude increases (see main Fig.\ \ref{fig04}), and is independent of the measurement location within the basin (see inset of Fig.\ \ref{fig04}), confirming the homogeneity of the wave field observed directly from the shore. 




\begin{figure}[!t]
\begin{center}
\includegraphics[scale=0.45]{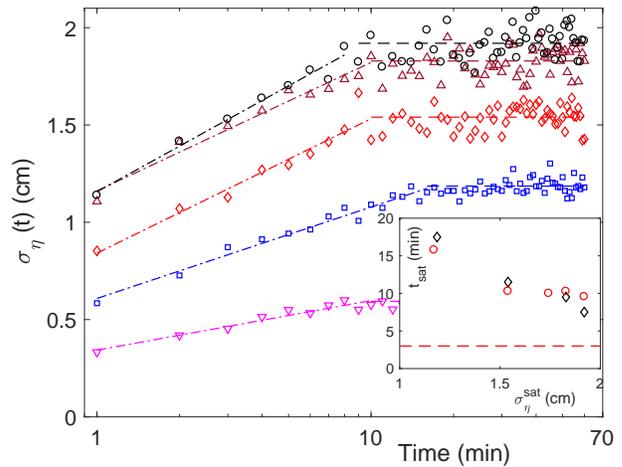}
\caption{Logarithmic time evolution of $\sigma_{\eta}(t)$ for different forcing amplitudes (same symbols as Fig.\ \ref{fig03}). Dotted-dash lines show logarithmic growth fits. Dashed lines shows the saturation values $\sigma^{sat}_{\eta}$. Inset: Saturation times of ($\circ$) $\sigma_{\eta}(t)$ and of ($\diamond$) the spectrum evolution (inferred from Fig.\ \ref{fig03}) vs forcing. Dashed line corresponds to the typical wave packet round-trip (3 min). $x=20$ m.} 
\label{fig05}
\end{center}
\end{figure}

To confirm that the inverse cascade dictates the saturation timescale of this system, we consider the temporal evolution of the standard deviation of the wave height $\sigma_{\eta}(t) = \sqrt {\int S_{\eta}(\omega,t) d \omega }$, again in the middle of the basin (see Fig.\ \ref{fig05}).  A  logarithmic growth $\sigma_{\eta}(t)=\gamma \log{(t/t_0)} +\sigma_{\eta}^{t_0}$ is clearly observed followed by a saturation to a value $\sigma_{\eta}^{sat}$ occurring at $t_{sat}$. The saturation time $t_{sat}=t_0\exp[(\sigma_{\eta}^{sat}-\sigma_{\eta}^{t_0})/\gamma]$ is found to be about 10 - 20 min ($\gamma \in [0.1, 0.4]$ cm and $t_0=1$ min), and to be consistent with the one related to $f_{sat}$ as displayed in Fig.\ \ref{fig05}. This demonstrates that, as could be inferred from the spectra reported in Fig.\ \ref{fig02}, a statistically steady-state is reached when the \textit{inverse} cascade saturates. Both estimates show that the saturation times are much more than a typical wave packet round-trip (3 min assuming 1D propagation), and decrease with the forcing amplitude, or with the wave steepness $\epsilon \equiv k_{max}\sigma^{sat}_{\eta} = (2\pi f^{sat}_{max})^2 \sigma^{sat}_{\eta}/g$, the right-hand axis of Fig.\ \ref{fig04} showing the relationship $\epsilon(\sigma^{sat}_{\eta})$.


So, what stops the inverse cascade? Since energy (and wave action) is continuously injected into this closed system, a steady state must involve either linear dissipation or nonlinear localized dissipative structures (e.g. sharp-crested waves, whitecapping \cite{Korotkevitch2008}, wave breakings, bound waves \cite{HerbertPRL10,Michel2018,CampagnePRF18}, or parasitic gravity-capillary waves \cite{FedorovPOF98}).
Prevailing linear dissipation at a wave frequency $f_0$ occurs in the viscous surface and basin lateral boundary layers yielding a typical decay time of $\tau_{diss} > 3000$ s \cite{Lamb}. This timescale is such that the scale separation required by WWT is verified, 
$\tau_{lin}(f)\ll \tau^ i_{nl}(f) \ll \tau_{diss}(f)$, between the linear propagation time $\tau_{lin}=1/\omega$, the nonlinear interaction time $\tau^i_{nl}$ and the dissipative time $\tau_{diss}$, for our range of $f$. Indeed, using the values of $Q$ and $P$ inferred experimentally (see below), and using the dimensionless constant value found experimentally for $\tau_{nl}$ \cite{DeikeJFM2015}, one finds $\tau^ i_{nl}$ at least two decades shorter (resp. longer) than $\tau_{diss}$ (resp. $\tau_{lin}$), with no fitting parameter (see Supp. Material \cite{SuppMat}). The typical discreteness $\tau_{disc}$ (i.e. inverse of the frequency separation of adjacent eigenmodes) is also found to be almost two decades longer than $\tau^ i_{nl}$ at the saturation frequency (see Supp. Material \cite{SuppMat}). Basin finite size effects are thus unlikely to affect the dynamics (see case n\textsuperscript{o}1 in \cite{Zakharov2005}). The saturation time (shorter than $\tau_{diss}$ and $\tau_{disc}$) depends on the wave steepness. It is thus due to a nonlinear effect, probably related to the sharp-crested waves, occurring homogeneously in the wave field, and visible directly from the shore once a steady state is reached (see movies in \cite{SuppMat}). Dissipation by nonlinear localized structures in the energy balance equation is indeed often referred in forecasting models of wind-driven ocean waves \cite{Badulin2005,KomenBook} and remains very challenging to estimate. Numerical simulations of fully nonlinear equations demonstrate that such structures are enhanced in the presence of an inverse cascade \cite{Korotkevitch2008}, and induce an effective large-scale dissipation not taken into account in WWT \cite{Zakharov2007,Korotkevich2019}. However, we have currently no way to quantify it since the wave probes are distributed discreetly over the basin surface, and localized structures are most of the time not captured. Indeed, the probability distributions of $\eta(t)$ and of $\partial \eta(t) / \partial t$ are found similar before and after the saturation, and close to a Tayfun distribution. Spatio-temporal measurements seem necessary to ascertain the role of localized structures, for instance by measuring the nonlinear corrections to the dispersion relation (see \cite{Korotkevich2013} for a numerical study), but are difficult to implement in such a large wave basin~\cite{Aubourg2017}.





\begin{figure}[!t]
\begin{center}
\includegraphics[scale=0.45]{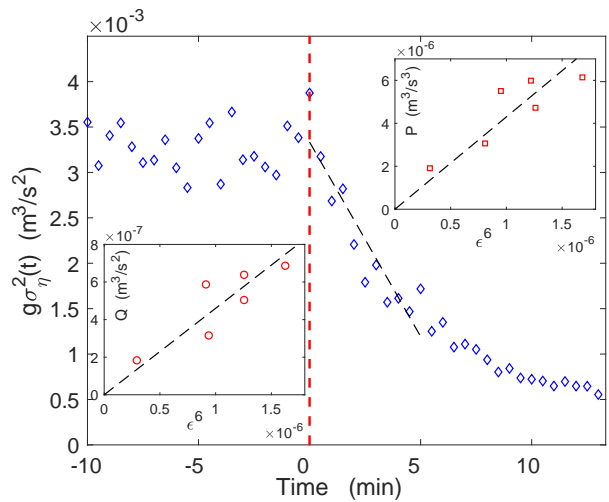}
\caption{Temporal evolution of the gravity wave energy. Wave maker is stopped at $t=0$. Dashed line: tangent at $t=0$ 
of slope $P=5.98\ 10^{-6}$ m$^3$/s$^3$. $\sigma^{sat}_{\eta}=1.82$ cm. $x=20$ m. Mean energy flux $P$ (top inset) and mean wave action flux $Q$ (bottom inset) vs. wave steepness to the power 6 from decaying experiments (main).} 
\label{fig06}
\end{center}
\end{figure}

\paragraph*{Wave action flux.\textemdash}One can finally estimate the mean cascading wave action flux $Q$ from the wave energy decay, notably to check further theoretical predictions of WWT. After 60 min of stationary wave turbulence, the wave maker is suddenly stopped at time $t=0$, and the temporal decay of the wave height is recorded during up to 30 min. Using the expressions of the gravity wave energy $E(t)=g\sigma^{2}_{\eta}(t)$ and of the power budget, 
the mean energy flux $P$ (per unit surface and density) is given by $P=-gd\sigma_{\eta}^2(t)/dt|_{t=0}$ \cite{DeikeJFM2015}. $P$ is thus experimentally estimated from the tangent at $t=0$ (see main Fig.\ \ref{fig06}), and the wave action flux from $Q \sim P/\omega_0$. We observe that $Q$ (and $P$) increases linearly with $\epsilon^6$ as expected theoretically \cite{Annenkov2006,Annenkov2009}. Beyond this agreement with WWT, $Q$ is found to be at least 4 orders of magnitude smaller than the critical flux breaking weak turbulence, $Q_c=g^2/\omega_0^4 \simeq 6\ 10^{-3}$ m$^3$s$^{-2}$, regardless $\epsilon$. 



\paragraph*{Conclusion.\textemdash}We reported the formation of an inverse cascade of gravity wave turbulence in a large-scale closed basin. Although it satisfies some assumptions of WWT (weak nonlinearity, timescale separation, scaling of $Q(\epsilon)$, and homogeneity), its time evolution is found to be stopped well before wavelengths sizable to the length of the basin are observed. This contrasts with the only laboratory observation performed so far in a small-scale basin where finite size-effects stops the inverse cascade~\cite{Deike2011}. Here, the saturation is related to a nonlinear effect probably the occurrence of highly nonlinear dissipative structures (sharp-crested waves) visible in the wave field. Our findings could be useful in different wave turbulence fields involving an inverse cascade such as Langmuir waves in plasmas \cite{Kats1971}, Kelvin waves in quantum fluid \cite{Ganshin2008}, or for elastic waves in thin plates where coherent structures were shown numerically to limit the inverse cascade dynamics \cite{During2015}. More generally, better identifying the mechanisms (such as inverse cascade) governing large-scale properties of turbulent flows is of paramount interest \cite{Alexakis2018}.



\begin{acknowledgments}
We thank S. Lambert, B. Pettinoti and J. Weis (ECN) for their technical help on the experimental setup. Part of this work was supported by the French National Research Agency (ANR DYSTURB project No. ANR-17-CE30-0004), and by a grant from the Simons Foundation MPS N$^{\rm o}$651463-Wave Turbulence. NM is supported by European Research Council (ERC N$^{\rm o}$647018-WATU).
\end{acknowledgments}
 

\end{document}